\documentclass[a4paper,11pt]{article}

\usepackage{lmodern}

\usepackage[utf8]{inputenc}
\usepackage[english]{babel}
\usepackage[myheadings]{fullpage}
\usepackage[T1]{fontenc}
\usepackage{subfig}
\usepackage{fancyhdr}
\usepackage{graphicx}
\usepackage{sectsty}
\usepackage{url}
\usepackage[usenames, dvipsnames, table, xcdraw]{xcolor}
\usepackage{amsmath}
\usepackage{amssymb}
\usepackage{calrsfs}
\usepackage{bbold}
\usepackage{dsfont}
\usepackage{esint}
\usepackage{indentfirst}
\usepackage{xfrac}
\usepackage{bigints}
\usepackage{slashed}
\usepackage{breqn}
\usepackage{autobreak}
\usepackage{setspace}
\usepackage[colorlinks]{hyperref}
\usepackage{microtype}
\usepackage{quoting}
\usepackage{bigints}
\usepackage{float}
\usepackage{array}
\usepackage[export]{adjustbox}
\usepackage[section]{placeins}
\usepackage[capitalise]{cleveref}
\usepackage{enumitem}
\usepackage{scalefnt}
\usepackage{booktabs}
\usepackage{caption}
\usepackage{makecell}
\usepackage{authblk}

\usepackage[normalem]{ulem}

\usepackage{cite}

\usepackage{xspace}

\usepackage{colortbl}
\definecolor{myblue}{rgb}{0.8,0.85,1}
\definecolor{light-gray}{gray}{0.95}

\hypersetup{colorlinks,
linkcolor={red!50!black},
citecolor={blue!50!black},
urlcolor={blue!80!black}}

\long \def \blockcomment #1\endcomment{}

\def\g{\gamma}

\def\m{\mu}

\def\p{\pi}

\def\r{\rho}
\def\s{\sigma}

\def\D{\Delta}

\def\half{\frac{1}{2}}

\def\units{\times 10^{-10}}

\makeatletter
\newcommand{\thickhline}{\noalign {\ifnum 0=`}\fi \hrule height 1pt
\futurelet \reserved@a \@xhline}
\newcolumntype{"}{@{\hskip\tabcolsep\vrule width 1pt\hskip\tabcolsep}}
\makeatother

\allowdisplaybreaks

\def\beq {\begin{equation}}
\def\eeq {\end{equation}}
\def\bea {\begin{eqnarray}}
\def\eea {\end{eqnarray}}

\def\nn {\nonumber}

\def\lbl{\label}

\def\beq{\begin{equation}}
\def\eeq{\end{equation}}
\def\bqry{\begin{eqnarray}}
\def\eqry{\end{eqnarray}}

\def\seEq#1{Eq.~(\ref{#1})}

\def\rcite#1{Ref.~\cite{#1}}

\def\amuHVP{a_\mu^{\rm HVP}}

\AtBeginDocument{}

\title{\LARGE {\bf \boldmath
Data-driven results for light-quark connected and strange-plus-disconnected
hadronic $g-2$ short- and long-distance windows \\ }}

\author[a]{Genessa Benton}
\author[b]{Diogo Boito}
\author[c]{Maarten Golterman}
\author[d]{Alexander Keshavarzi}
\author[e,f]{Kim Maltman}
\author[g]{Santiago~Peris\vspace{0.5cm}}

\affil[a]{\it Department of Physics, University of Illinois, Urbana, IL 61801, USA\vspace{0.2cm}}

\affil[b]{\it Instituto de F\'isica de S\~ao Carlos, Universidade de S\~ao Paulo, CP 369, 13560-970, S\~ao Carlos, SP, Brazil\vspace{0.2cm}}

\affil[c]{\it Department of Physics and Astronomy, San Francisco State University,
San Francisco, CA 94132, USA\vspace{0.2cm}}

\affil[d]{\it Department of Physics and Astronomy, The University of Manchester,
Manchester M13 9PL, United Kingdom\vspace{0.2cm}}

\affil[e]{\it Department of Mathematics and Statistics,
York University, Toronto, ON Canada M3J~1P3\vspace{0.2cm}}

\affil[f]{\it CSSM, University of Adelaide, Adelaide, SA~5005 Australia\vspace{0.2cm}}

\affil[g]{\it Department of Physics and IFAE-BIST, Universitat Aut\`onoma de Barcelona,
E-08193 Bellaterra, Barcelona, Spain
\vspace{0.3cm}}

\date{}

\begin{document}

\begin{flushright}
{\small \today}
\end{flushright}

\vspace*{-0.7cm}
\begingroup
\let\newpage\relax
\maketitle
\endgroup

\vspace*{-1.0cm}
\begin{abstract}
\noindent
A key issue affecting the attempt to reduce the
uncertainty on the Standard Model prediction for the muon anomalous
magnetic moment is the current discrepancy between lattice-QCD
and data-driven results for the hadronic vacuum polarization.
Progress on this issue
benefits from precise
data-driven determinations of the isospin-limit light-quark-connected
(lqc) and strange-plus-light-quark-disconnected (s+lqd) components of
the related RBC/UKQCD windows. In this paper, using a strategy
employed previously for the intermediate window, we provide
data-driven results for the lqc and s+lqd components of the short-
and long-distance RBC/UKQCD windows. Comparing these results
with those from the lattice, we find significant discrepancies
in the lqc parts but good agreement for the s+lqd components.
We also explore the impact of recent CMD-3
$e^+e^-\to \pi^+\pi^-$ cross-section results, demonstrating that
an upward shift in the $\rho$-peak region of the type seen in the
CMD-3 data serves to eliminate the discrepancies for the
lqc components without compromising the good agreement between
lattice and data-driven s+lqd results.

\end{abstract}

\thispagestyle{empty}

\clearpage
\vspace*{0.0cm}

\setcounter{page}{1}
\section{Introduction}

In 2021 and 2023, the Fermilab E989 experiment published new
measurements~\cite{Muong-2:2021ojo,Muong-2:2021vma,Muong-2:2023cdq}
of the anomalous magnetic moment of the muon, $a_\mu$. The new results
are compatible with the older, less precise, BNL E821
determination~\cite{Muong-2:2006rrc} and produce a new
experimental $a_\mu$ world average with an impressive precision of
0.19 ppm. Prior to the release of the E989 results, in 2020,
the Muon $g-2$ Theory Initiative published a White
Paper~\cite{Aoyama:2020ynm} giving the then-best Standard Model
(SM) prediction for $a_\mu$, based on the results of
Refs.~\cite{Aoyama:2012wk,Aoyama:2019ryr,Czarnecki:2002nt,Gnendiger:2013pva,Davier:2017zfy,Keshavarzi:2018mgv,Davier:2019can,Keshavarzi:2019abf,Colangelo:2018mtw,Hoferichter:2019mqg,Hoid:2020xjs,Kurz:2014wya,Melnikov:2003xd,Masjuan:2017tvw,Colangelo:2017qdm,Colangelo:2017fiz,Hoferichter:2018dmo,Hoferichter:2018kwz,Gerardin:2019vio,Bijnens:2019ghy,Colangelo:2019lpu,Colangelo:2019uex,Colangelo:2014qya,Blum:2019ugy}.
This prediction, which employed the then-current data-driven
result for the hadronic vacuum polarization (HVP) contribution,
was in disagreement with the new experimental result.

Since the 2020 White Paper, three important developments have taken
place. First, in 2021, the BMW collaboration published a complete and
competitive lattice-QCD determination of the HVP contribution,
$a_\mu^{\rm HVP}$, to $a_\mu$~\cite{Borsanyi:2020mff}. The
modified SM prediction produced by this result is in agreement with
experiment within less than $2\sigma$.\footnote{An early sign
of discrepancy between lattice QCD and the data-driven approach was the
light-quark connected RBC/UKQCD intermediate-window lattice result
of Ref.~\cite{Aubin:2019usy}, which was significantly larger than
a data-driven estimate.}
Second, new experimental
results for the cross section of $e^+e^-\to \pi^+\pi^-$ obtained by
the CMD-3 experiment~\cite{CMD-3:2023alj} were found to produce a
$2\pi$ contribution to $\amuHVP$ significantly larger than that implied
by previous experiments~\cite{Aoyama:2020ynm}, raising further
questions about the data-driven evaluation. Finally, very recently, in
Ref.~\cite{Boccaletti:2024guq}, a more precise determination of $\amuHVP$
was obtained using new BMW lattice-QCD results combined with a data-driven
evaluation, based on Ref.~\cite{Davier:2019can}, of the long-distance
contribution from a region in Euclidean time where the lattice-QCD
determination is significantly less precise. This leads to a result
for $a_\mu$ that agrees with experiment to within less than $1\sigma$.

Given this situation, it has become critical to understand the present
discrepancy between the data-driven and lattice-QCD results for
$a_\mu^{\rm HVP}$ in more detail. An important tool for the comparison
between lattice-QCD and data-driven results is the method of
``windows,'' introduced by RBC/UKQCD~\cite{RBC:2018dos}. The
method involves splitting the $\amuHVP$ integral in three
parts with a short-, an intermediate-, and a long-distance
contribution. For the intermediate window, which significantly
suppresses lattice uncertainties associated with the continuum limit
and finite volume effects, and which can be computed with very good
statistical precision, it has been shown that the data-driven approach
and the lattice-QCD results display a significant tension, of about
$4\sigma$~\cite{Colangelo:2022vok,Wittig:2023pcl}.\footnote{This
result predates the publication of the new CMD-3 measurement for
$e^+e^-\to \pi^+\pi^-$~\cite{CMD-3:2023alj} mentioned above.}$^{,}$\footnote{With $\tau$ data for the two-pion contribution,
together with the estimate for the
required isospin-breaking corrections obtained from the model of Ref.~\cite{Miranda:2020wdg}, this discrepancy
is reduced~\cite{Masjuan:2023qsp}.}
Since the lattice intermediate window result is the sum of a number
of different components, it is of interest to further refine this
comparison in order to clarify the origin of the observed discrepancy.

The ingredients that go into the lattice determination are as follows.
The main contribution arises from the isospin-symmetric light-quark
connected diagrams, followed in size by the strange-quark connected
and three-flavor disconnected contributions. Smaller charm-
and bottom-quark contributions (both connected and disconnected) are
also included. Electromagnetic (EM) and strong-isospin-breaking (SIB)
corrections are treated perturbatively to first order in the fine-structure
constant $\alpha$ and the up-down quark-mass difference, which is
sufficient for the present precision. The data-driven approach, in
turn, is based on the analysis of exclusive hadronic electroproduction
data, channel by channel, up to squared hadronic invariant
masses, $s$, of about $4$~GeV$^2$, and inclusive data
and/or perturbative QCD (pQCD) at higher $s$.

Several lattice collaborations have produced partial results for the
different contributions to $\amuHVP$ using the RBC/UKQCD and other
windows. For the intermediate window, ten recent, mutually compatible,
lattice-QCD results for the isospin-symmetric light-quark connected
(lqc) component now exist~\cite{Borsanyi:2020mff,Lehner:2020crt,Wang:2022lkq,Aubin:2022hgm,Ce:2022kxy,ExtendedTwistedMass:2022jpw,FermilabLatticeHPQCD:2023jof,RBC:2023pvn,Boccaletti:2024guq,Bazavov:2024xsk}.
Results for the lqc short-distance contribution have also been obtained by
six independent groups~\cite{Kuberski:2024bcj,Boccaletti:2024guq,RBC:2023pvn,ExtendedTwistedMass:2022jpw,Bazavov:2024xsk,Spiegel:2024dec} and, recently, two results for the lqc contribution to the long-distance
window have been published~\cite{RBC:2024fic,Djukanovic:2024cmq}.
Other sub-dominant contributions, such as the full strange-quark and
light-quark disconnected contributions have also been calculated by
different lattice-QCD
collaborations~\cite{Boccaletti:2024guq,Kuberski:2024bcj,Borsanyi:2020mff,RBC:2023pvn,ExtendedTwistedMass:2022jpw,FermilabLatticeHPQCD:2023jof,Gerardin:2019rua,Budapest-Marseille-Wuppertal:2017okr,Blum:2015you,RBC:2018dos}.

In order to sharpen the comparison of the data-driven and lattice-QCD
approaches to $\amuHVP$ it is highly desirable to have data-driven estimates
of the different components of the lattice results. In recent
papers~\cite{Boito:2022rkw,Boito:2022dry,Benton:2023dci,Benton:2023fcv},
we have shown that it is possible to reorganize the data-driven
computation of $\amuHVP$ to reliably estimate the lqc, as well as
the complete strange-quark plus light-quark disconnected
(s+lqd), contributions to $\amuHVP$. We demonstrated that the discrepancy
with pre-CMD-3 data-driven results~\cite{Keshavarzi:2019abf,Davier:2019can}
originates mostly in the intermediate window isospin-symmetric lqc
contribution~\cite{Benton:2023dci} while s+lqd data-driven results are
in good agreement with their lattice counterparts~\cite{Benton:2023fcv}.
In the present paper, we obtain the data-driven determinations of the
lqc and s+lqd components of the short- and long-distance
isospin-symmetric RBC/UKQCD windows. We present results based on
the pre-CMD-3 data combination of Ref.~\cite{Keshavarzi:2019abf},
referred to as the ``KNT19 data,''
but estimate, as well, the potential impact
of the new CMD-3 results for the $\pi^+\pi^-$ electroproduction
cross sections. We emphasize that this latter study, using
CMD-3 data, represents only a first exploration, since, given the
present discrepancies in the $2\pi$ cross-section data-base, a meaningful
combination of all experimental $2\pi$ data appears impossible. This
paper concludes our data-driven determinations of the lqc and s+lqd
contributions to the RBC/UKQCD windows and provides additional
benchmarks for the comparison with present and future lattice-QCD
results.

The paper is organized as follows. In Sec.~\ref{sec:review}, we
define the RBC/UKQCD windows that are studied in this paper and
briefly review the strategy for the data-driven determination of
the lqc and s+lqd contributions to $\amuHVP$ and the intermediate
RBC/UKQCD window already employed in
Refs.~\cite{Boito:2022rkw,Boito:2022dry,Benton:2023dci,Benton:2023fcv}.
In Sec.~\ref{sec:implementation}, we discuss the implementation of this
strategy for the short-distance (SD) and long-distance (LD) windows
and give intermediate results for the different contributions to the lqc
and s+lqd components of the windows. In Sec.~\ref{sec:final-results}, we
present our final, pre-CMD-3 data-based results for the SD and
LD windows, quoting, for completeness, also the results for the
intermediate window and the total HVP contribution, already presented in
Refs.~\cite{Boito:2022rkw,Boito:2022dry,Benton:2023dci,Benton:2023fcv}.
In Sec.~\ref{sec:CMD-3-results} we make an exploratory study of the
potential impact of the CMD-3 results for the $2\pi$ channel. Our
conclusions are presented in Sec.~\ref{sec:conclusions}.

\section{Review of the general strategy}
\label{sec:review}

Here, we present a concise review of the strategy employed in this
paper. It has been described in detail in our previous
publications~\cite{Boito:2022rkw,Boito:2022dry,Benton:2023dci,Benton:2023fcv}
to which we refer for additional details.

\subsection{Windows}
We start by defining the window quantities considered in this paper.
The master formula for the data-driven (or dispersive) approach to the
leading order $\amuHVP$ is~\cite{Brodsky:1967sr,Lautrup:1968tdb,Gourdin:1969dm}
\beq
\lbl{amuHVP}
a_\m^{\rm HVP}=\frac{4\alpha^2m_\mu^2}{3}
\int_{m_\pi^2}^\infty ds\, \frac{\hat{K}(s)}{s^2} \r_{\rm EM}(s),
\eeq
where $m_\p$ is the neutral pion mass and $\r_{\rm EM}(s)$ represents the
inclusive EM-current hadronic spectral function, which is related to the
$R$ ratio derived from the bare inclusive hadronic electroproduction
cross section, $\s^{(0)}[e^+ e^-\rightarrow \mbox{hadrons}(+\g)]$, by
\bqry
\lbl{rdefn}
\r_{\rm EM}(s)&=&\frac{1}{12\p^2}R(s),\\
R(s)&=&\frac{3s}{4\pi \alpha}\,
\sigma^{(0)} [e^+ e^-\rightarrow \mbox{hadrons}(+\gamma)],\nonumber
\eqry
where the kernel $\hat{K}(s)$ is a smoothly varying, monotonically
increasing function with $\hat{K}(4m_\p^2)\approx 0.63$ at the two-pion
threshold and $\lim_{s\to\infty}\hat{K}(s)=1$.\footnote{For the explicit
expressions, we refer to e.g.~\rcite{Aoyama:2020ynm}.} In terms of the
so-called time-momentum representation, $\amuHVP$ can be obtained from
the Euclidean-time two-point correlator~\cite{Bernecker:2011gh}
\beq
\lbl{defC}
C(t)=\frac{1}{3}\sum_{i=1}^3\int d^3x \langle j_i^{\rm EM}(\vec{x},t)
j_i^{\rm EM}(0)\rangle=\half\int_{m_\p^2}^\infty ds\,\sqrt{s}\,
e^{-\sqrt{s}t}\,\r_{\rm EM}(s)\quad (t>0),
\eeq
built from the EM current $j_i^{\rm EM}(\vec x,t)$ as
\beq
\lbl{amuHVPtimemom}
a_\m^{\rm HVP}=2\int_0^\infty dt\,w(t) C(t),
\eeq
where the function $w(t)$ is known and can be obtained from
$\hat K(s)$.\footnote{For explicit expressions and a useful
approximation see Refs.~\cite{Bernecker:2011gh,DellaMorte:2017dyu}.}

The windowed versions (``windows'' in what follows) we consider
in this work, originally introduced in~\rcite{RBC:2018dos}, are the
SD, intermediate (int), and LD windows obtained by inserting the
functions
\begin{align}
\lbl{windows}
W_{\rm SD}(t)&=1-\Theta(t,t_0,\D), \nonumber \\
W_{\rm int}(t)&=\Theta(t,t_0,\D)-\Theta(t,t_1,\D), \nonumber \\
W_{\rm LD}(t)&=\Theta(t,t_1,\D),
\end{align}
respectively, with
\beq
\Theta(t,t^\prime, \Delta) = \half\left( 1+\tanh\frac{t-t^\prime}{\D} \right)
\eeq
and $t_0=0.4\,{\rm fm}$, $t_1=1.0\,{\rm fm}$, and $\D=0.15\,{\rm fm}$,
into the representation of \seEq{amuHVPtimemom}, leading~to
\bqry
\lbl{amuW}
a_\m^{\rm win}&=&2\int_0^\infty dt\,W_{\rm win}(t)\,w(t) C(t) \\
&=& \frac{4\alpha^2m_\mu^2}{3}\int_{m_\p^2}^\infty ds\,
\frac{\hat{K}(s)}{s^2} \,\widetilde{W}_{\rm win}(s)\,\r_{\rm EM}(s),\nonumber
\eqry
where ${\rm win}=\{ {\rm SD},{\rm int}, {\rm LD}\}$ and the window
function in $s$-space is written as
\beq
\lbl{windowint}
\widetilde{W}_{\rm win}(s)=\frac{\int_0^\infty dt\,W_{\rm win}(t)\,
w(t)\,e^{-\sqrt{s}t}}{\int_0^\infty dt\,w(t)\,e^{-\sqrt{s}t}}.
\eeq
By construction, $\amuHVP$ is given by the sum of the three windows
defined above.

\subsection{Light-quark connected and strange plus disconnected spectral
functions in the isospin limit}

Our data-driven determination of the lqc and the s+lqd contributions
to $\amuHVP$ follows the strategy implemented for the first time in
Refs.~\cite{Boito:2022rkw,Boito:2022dry}. We start from the decomposition
of the three-light-flavor EM current into its $I=1$ (flavor
octet label 3) and $I=0$ (flavor octet label 8) parts, and the
related decompositions of $\rho_{\rm EM}(s)$, $C(t)$ and any weighted
integrals thereof, into their pure $I=1/0$ (flavor octet labels 33/88) and
mixed-isospin (MI, flavor octet labels 38) components. In the
isospin limit, the pure $I=1$ contributions are entirely
light-quark connected and the corresponding pure $I=0$ lqc
contributions exactly 1/9 times their $I=1$ lqc counterparts. Thus,
for example, the total lqc contribution to the EM spectral
function is
\beq
\rho^{\rm lqc}_{\rm EM}(s) = \frac{10}{9}\rho_{\rm EM}^{I=1}(s),
\label{eq:rholqc}
\eeq
and one can obtain the full strange plus light-quark
disconnected contribution from the combination
\begin{align}
\rho^{\rm s+lqd}_{\rm EM}(s) &= \rho_{\rm EM}^{I=0}(s)
-\frac{1}{9}\rho_{\rm EM}^{I=1}(s) \nonumber \\
&=\rho_{\rm EM}(s)-\frac{10}{9}\rho_{\rm EM}^{I=1}(s). \label{eq:rhosplqd}
\end{align}
Our main task is, hence, the identification, with sufficient precision,
of the $I=1$ and $I=0$ components of $\rho_{\rm EM}$ and any
associated weighted integral quantities. As outlined below, this can be
accomplished on a channel-by-channel basis, using isospin symmetry,
in the exclusive-mode region of the $R(s)$ data. In this work, we employ
the results for the exclusive-mode components of the pre-CMD-3
$R(s)$ data combination of Ref.~\cite{Keshavarzi:2019abf}, commonly
referred to as `KNT19.' It would be interesting to perform our
analysis using other data combinations, such as that of
\rcite{Davier:2019can}, but this would require access to their exclusive
spectra, channel-by-channel and with all correlations, which, to the
best of our knowledge, is not publicly available.

Substituting the spectral functions for the lqc and s+lqd components,
Eq.~(\ref{eq:rholqc}) and Eq.~(\ref{eq:rhosplqd}), respectively,
into Eq.~(\ref{amuW}), one obtains the respective isospin-limit contributions
to each of the $a_\mu^{\rm HVP}$ windows, which we denote by
$a_\mu^{\rm win, lqc}$ and $a_\mu^{\rm win, s+lqd}$.

\section{Implementation and intermediate results}
\label{sec:implementation}

Four different ingredients are required for our determinations of the
isospin-symmetric lqc and s+lqd components, $a_\mu^{\rm win, lqc}$ and
$a_\mu^{\rm win, s+lqd}$, of the RBC/UKQCD windows defined in
Eqs.~(\ref{windows}) to (\ref{amuW}). The first is the set of contributions
from exclusive modes which are $G$-parity eigenstates, where the $G$-parity
allows the isospin of the contribution to be unambiguously identified. We
refer to such modes generically as ``unambiguous.'' The sum of
$G$-parity-positive-mode contributions dominates the lqc component of the
three windows. We discuss the unambiguous-mode contributions in
Sec.~\ref{sec:unamb}. The second ingredient is the set of $I=0$ and $1$
components of contributions from exclusive modes that are not $G$-parity
eigenstates, which we refer to as ``ambiguous''. The isospin separation
for such ambiguous-mode contributions is discussed in Sec.~\ref{sec:amb}.
For the KNT19 compilation, the ``exclusive-mode region'' (the region of
squared hadronic invariant masses, $s$, in which $R(s)$ is saturated by
the sum of measured exclusive-mode contributions) extends up to
$s=(1.937~{\rm GeV})^2$. We refer to the region above this as the
``inclusive region.'' The third ingredient
is the set of inclusive-region contributions. For these we employ
(pQCD), supplemented with an estimate for the impact
of quark-hadron duality violation, as explained in Sec.~\ref{sec:pert}.
The fourth ingredient, discussed in Sec.~\ref{sec:IB},
is the EM and SIB corrections that must be applied before
performing comparisons with isospin-symmetric lattice-QCD
results.

\subsection{Modes with unambiguous isospin}
\label{sec:unamb}

As discussed in detail in
Refs.~\cite{Boito:2022rkw,Boito:2022dry,Benton:2023dci,Benton:2023fcv},
in the isospin limit, modes with positive (negative) $G$-parity
have isospin $I=1$ ($I=0$). In Tabs.~\ref{tab:unambiguous-modes-I1}
and~\ref{tab:unambiguous-modes-I0}, we give the contributions of each of
the $G$-parity unambiguous modes $X$ present in the KNT19 data combination
to the three windows of Eq.~(\ref{windows}), denoted
$[a_\mu^{\rm win}]_X$, as well as to the total HVP contribution,
$[a_\mu^{\rm HVP}]_X$.{\footnote{Entries whose mode names include
the phrase ``low-$s$'' are those labelling very-near-threshold contributions
for which the underlying $R(s)$ contributions were obtained using Chiral
Perturbation Theory in the KNT19 exclusive-mode compilation.} Although
the central value for $[a_\mu^{\rm HVP}]_X$ is always the sum of
the central values of the contributions to the three windows from
the same mode, the uncertainty in the fifth column of
Tabs.~\ref{tab:unambiguous-modes-I1} and ~\ref{tab:unambiguous-modes-I0}
includes the effect of correlations
among the errors on the three associated window quantities.
The results for the
intermediate window and for the total HVP were already given in
Refs.~\cite{Benton:2023dci,Boito:2022rkw} and are repeated here for completeness.

\begin{table}[!t]
\begin{center}
\caption{\label{tab:unambiguous-modes-I1}
Contributions from $G$-parity positive modes (hence $I=1$) to
$a_\mu^{\rm SD}$, $a_\mu^{\rm int}$, $a_\mu^{\rm LD}$, and
$a_\mu^{\rm tot}$ for $\sqrt{s}\leq 1.937$~GeV obtained from
KNT19~\cite{Keshavarzi:2019abf} exclusive-mode spectra. The
label ``npp'' stands for ``non-purely pionic.'' All entries in
units of $10^{-10}$. Uncertainties in the last column take
into account the correlations between the three windows.}
\begin{tabular}{lllll}
\toprule
$I=1$ modes $X$&$[a_\mu^{\rm SD}]_X\times 10^{10}$ &
$[a_\mu^{\rm int}]_X\times 10^{10}$ & $[a_\mu^{\rm LD}]_X\times 10^{10}$
&$[a_\mu^{\rm HVP}]_X\times 10^{10}$ \\
\hline
low-$s$ $\pi^+ \pi^-$ & 0.0010(00) & 0.02(00) & 0.842(18)
& 0.867(18)\\
$\pi^+ \pi^-$ & 14.927(52) & 144.13(49) & 344.4(1.4)
& 503.5(1.9)\\
$2\pi^+ 2\pi^-$ & 3.239(43) & 9.29(13) & 2.334(34)
& 14.87(20)\\
$\pi^+ \pi^- 2\pi^0$ & 3.98(16) & 11.94(48) & 3.46(14)
& 19.39(78) \\
$3\pi^+ 3\pi^-$ (no $\omega$) & 0.0746(47)& 0.14(01) & 0.01485(95)
& 0.231(15)\\
$2\pi^+2\pi^-2\pi^0$ (no $\eta$) & 0.426(52) & 0.83(11) & 0.092(13)
& 1.35(17)\\
$\pi^+\pi^- 4\pi^0$ (no $\eta$) & 0.067(67) & 0.13(13) & 0.014(14)
& 0.21(21)\\
$\eta \pi^+ \pi^-$ & 0.333(12) & 0.85(03) & 0.1594(63)
& 1.340(50)\\
$\eta 2\pi^+ 2\pi^-$ & 0.0239(33) & 0.05(01) & 0.00547(98)
& 0.076(11)\\
$\eta \pi^+\pi^- 2\pi^0$ & 0.0407(66) & 0.07(01) & 0.0065(11)
& 0.119(20)\\
$\omega (\rightarrow \pi^0\gamma)\pi^0$ & 0.1469(34) & 0.53(01) & 0.2014(42)
& 0.882(19)\\
$\omega (\rightarrow {\rm npp})3\pi$ & 0.0529(99) & 0.10(02)& 0.0116(23)
& 0.168(32)\\
$\omega \eta \pi^0$ & 0.081(18) & 0.15(03) & 0.0144(29)
& 0.242(53)\\
\hline
Total ($I=1$) & 23.40(19) & 168.24(72) & 351.6(1.4) & 543.2(2.1)\\
\bottomrule
\end{tabular}
\end{center}
\end{table}

The corresponding exclusive-mode contributions to the full lqc
window totals are obtained by multiplying the entries in
Tab.~\ref{tab:unambiguous-modes-I1} by 10/9, as per Eq.~(\ref{eq:rholqc}).
The resulting sums of lqc contributions from all unambiguous
modes are listed in Tab.~\ref{tab:final-results-lqc}. The
contributions of the unambiguous modes to the s+lqd components
are given by the combination shown in Eq.~(\ref{eq:rhosplqd});
the totals of these contributions for each window appear in
Tab.~\ref{tab:final-results-slqd}.

\begin{table}[!t]
\begin{center}
\caption{\label{tab:unambiguous-modes-I0}
Contributions from $G$-parity negative modes (hence $I=0$) to
$a_\mu^{\rm SD}$, $a_\mu^{\rm int}$, $a_\mu^{\rm LD}$, and
$a_\mu^{\rm HVP}$ for $\sqrt{s}\leq 1.937$~GeV obtained from
KNT19~\cite{Keshavarzi:2019abf} exclusive-mode spectra. The label
``npp'' stands for ``non-purely pionic.''
Uncertainties in the last column take into account the correlations
between the three windows. All entries in units of $10^{-10}$.}
\begin{tabular}{lllll}
\toprule
$I=0$ modes $X$&$[a_\mu^{\rm int}]_X\times 10^{10}$
& $[a_\mu^{\rm SD}]_X\times 10^{10}$ & $[a_\mu^{\rm LD}]_X\times 10^{10}$
&$[a_\mu^{\rm tot}]_X\times 10^{10}$ \\
\hline
low-$s$ $3\pi$ & 0.0003 & 0.003
& 0.01 & 0.014 \\
$3\pi$ & 2.609(46)
& 18.69(35) & 25.42(54) & 46.73(94) \\
$2\pi^+2\pi^-\pi^0$ (no $\omega$, $\eta$) & 0.266(24)
& 0.613(57) & 0.100(10) & 0.979(90)\\
$\pi^+\pi^-3\pi^0$ (no $\eta$) & 0.172(30)
& 0.388(72) & 0.059(12) & 0.62(11)\\
$3\pi^+3\pi^-\pi^0$ (no $\omega$, $\eta$) & $-0.0010(17)$
& $-0.0027(32)$ & $-0.00053(35)$ & $-0.0043(53)$\\
$\eta\pi^+\pi^-\pi^0$ (no $\omega$) & 0.209(23)
& 0.441(51) & 0.0566(72) & 0.706(81)\\
$\eta\omega$ & 0.0823(62)
& 0.187(14) & 0.0267(20) & 0.296(22)\\
$\omega(\to {\rm npp})2\pi$ & 0.0370(40) & 0.0836(92)
& 0.0124(14) & 0.133(15)\\
$\omega 2\pi^+2\pi^-$ & 0.00246(69)
& 0.0045(13) & 0.00044(13) & 0.0074(21)\\
$\eta\phi$ & 0.1225(57)
& 0.253(0.012) & 0.0304(15) & 0.406(19)\\
$\phi\to({\rm unaccounted})$ & 0.0036(36)
& 0.02(0.02) & 0.017(17) & 0.043(43)\\
\hline
Total ($I=0$) & 3.502(65) & 20.69(37) & 25.74(54) & 49.93(95) \\
\bottomrule
\end{tabular}
\end{center}
\end{table}

\subsection{Modes with ambiguous isospin}
\label{sec:amb}

We turn now to the modes with no definite isospin. These are
of two distinct types, those for which external information can be
used to help separate the different isospin components and those
for which this is not possible.
Fortunately such external
information is available for the channels, $K\bar K$, $K\bar K \pi$
and $\pi^0/\eta +\gamma$, which give the largest of the ambiguous-mode
contributions. Contributions from other ambiguous modes all turn out to
be small. The strategy used to perform the ambiguous-mode isospin
separations is described in Ref.~\cite{Benton:2023fcv} (see
also \rcite{Boito:2022rkw}) and reviewed briefly below.

We start from those ambiguous modes having only $I=0$ and $I=1$
contributions in the isospin limit and for which no external information
is available. The small contributions from these modes are treated
using a ``maximally conservative'' prescription, based on the observation
that, because of spectral positivity, the $I=0$ and $I=1$ parts of the
contribution from the given mode $X$ must both lie between 0 and the full
experimental $I=1+0$ total obtained from the KNT19 $R(s)$ data. The
$I=0$ and $I=1$ components are then guaranteed to lie, respectively,
in the ranges ($50\pm 50$)\% and ($50\mp 50$)\% times the $I=0+1$ total,
with the two errors 100\% anticorrelated. The lqc and the s+lqd
parts of the mode-$X$ contribution to the EM spectral function then
lie in the following ranges~\cite{Boito:2022rkw}
\begin{align}
[\rho_{\rm EM}^{\rm lqc}]_X &= \left( \frac{5}{9}\pm\frac{5}{9}
\right)[\rho_{\rm EM}]_X, \nonumber \\
[\rho_{\rm EM}^{\rm s+lqd}]_X &= \left( \frac{4}{9}\pm\frac{5}{9}
\right)[\rho_{\rm EM}]_X.\label{eq:max-conservative}
\end{align}
This maximally conservative separation of $[\rho_{\rm EM}]_X$ produces
related results for the contributions to the window quantities
$[a_\mu^{\rm win, lqc}]_X$ and $[a_\mu^{\rm win,s+lqd}]_X$ when
used in Eq.~(\ref{amuW}).

It is crucial, however, to have better control over the dominant
contributions from ambiguous modes than would be provided by the maximally
conservative treatment, especially those arising from the
$K\bar K$ and, to a lesser extent, the $K\bar K\pi$ channels. For
these channels, external experimental information can be used to assess
the $I=1$ component of the total $I=1+0$ sum. For the $K\bar K$
modes ($K^+K^-$ and $K^0\bar K^0$), this external information
comes in the form of BaBar's measurement of the differential decay
distribution of $\tau \to K^-K^0\nu_\tau$~\cite{BaBar:2018qry}, which
provides a determination of the $K\bar{K}$ contribution to the
charged-current $I=1$ vector spectral function, which in turn,
via the conserved vector current (CVC) relation, provides a
determination of the $I=1$ part of the $K\bar{K}$ contribution to
$\rho_{EM}(s)$, and hence\footnote{For our purposes, isospin-breaking
corrections to the CVC relation can safely be neglected. In the case of
Eq.~(\ref{eq:KK-LD-lqc}), e.g., a 1\% isospin-breaking correction
would amount to $0.002\units$, which can very safely be neglected
when compared with other uncertainties entering this determination.}
an estimate of the $I=1$ part of the $K\bar{K}$ contribution to the
various window integrals in the region up to the endpoint,
$s = 2.7556~{\rm GeV}^2$, of the BaBar $\tau$ data.
Above this point, we integrate the $I=1+0$ tail of the KNT19 $K\bar{K}$
spectrum, applying the maximally conservative separation to this
small remainder to obtain our final $I=1$ and $I=0$ totals for the
full exclusive-region $K\bar K$ contributions. As an example, for
the lqc contribution to the LD window, this procedure gives, using
Eq.~(\ref{eq:rholqc}),
\beq
[a_\mu^{\rm LD}]^{\rm lqc}_{K\bar K} = \frac{10}{9}(0.1743(84)+0.0065(65))
\times 10^{-10} = 0.201(12)\times 10^{-10},\label{eq:KK-LD-lqc}
\eeq
where the first number in parenthesis is the $I=1$ contribution obtained
using BaBar $\tau \to K^-K^0\nu_\tau$ data and the second is the
contribution from the tail of the KNT19 spectrum, evaluated using
the maximally conservative separation of Eq.~(\ref{eq:max-conservative}).
Results for the other windows are given in the second row of
Tab.~\ref{tab:all-amb-lqc}. The s+lqd LD window $K\bar{K}$
contribution is, similarly, using Eq.~(\ref{eq:rhosplqd}),
\[
[a_\mu^{\rm LD}]^{\rm s+lqd}_{K\bar K} = \left[13.37(11)
-\frac{1}{9}\times 0.181(11)\right]\times 10^{-10} = 13.35(11)\times 10^{-10},
\]
where the first number in square brackets is the $I=0$ total and the
second number (one-ninth of) the $I=1$ part. The dominance of the $I=0$
component is a result of the enhancement produced by the $\phi$
resonance. The analogous results for the other windows are given
in the first row of Tab.~\ref{tab:all-amb-splqd}.

\begin{table}[!t]
\begin{center}
\caption{\label{tab:all-amb-lqc}
Contributions of modes with no definite isospin to the lqc parts of
$a_\mu^{\rm SD}$, $a_\mu^{\rm int}$, $a_\mu^{\rm LD}$, and
$a_\mu^{\rm HVP}$. See the text for the details of the treatment
of each channel. The label ``npp'' stands for ``non-purely pionic.''
Uncertainties in the last column take into account the correlations
between the three windows. All entries in units of $10^{-10}$. }
{\small
\begin{tabular}{lllll}
\toprule
mode &$[a_\mu^{\rm SD}]_X^{\rm lqc}\times 10^{10}$
& $[a_\mu^{\rm int}]_X^{\rm lqc}\times 10^{10}$
& $[a_\mu^{\rm LD}]_X^{\rm lqc}\times 10^{10}$
&$[a_\mu^{\rm tot}]_X^{\rm lqc}\times 10^{10}$ \\
\hline
$K\bar K$ & 0.167(30) & 0.579(64)
& 0.201(12) & 0.95(10) \\
$K\bar K \pi$ & 0.219(36) &0.521(86)
& 0.083(14) & 0.82(14) \\
$\pi^0\gamma +\eta\gamma$ & 0.0165(15) & 0.137(13)
& 0.204(21) & 0.36(4) \\
$K\bar K 2\pi$ & 0.32(32) &0.60(60)
& 0.062(62) & 0.98(98) \\
$K\bar K 3\pi$ & 0.0082(82) &0.012(12)
& 0.00054(54)& 0.21(21) \\
low-$s$ $\pi^0\gamma + \eta \gamma$ &0.00054(54) &0.0082(82)
& 0.061(61) & 0.070(70) \\
$N\bar N$ & 0.012(12) &0.019(19)
& 0.0015(15) & 0.033(33) \\
$\eta (\to{\rm npp}) K\bar K ({{\rm no}\, \phi})$ & 0.0027(27)
&0.0050(50) & 0.00052(52) & 0.0082(82) \\
$\omega (\to{\rm npp})$ $K\bar K$ & 0.00085(85)
&0.0012(12) & 0.000056(56) & 0.0021(21) \\
\hline
Total (lqc) & 0.75(32) & 1.88(61) & 0.613(91)
& 3.2(1.0) \\
\bottomrule
\end{tabular}}
\end{center}
\end{table}

For the $K\bar K \pi$ contributions, the needed external information
is provided by BaBar's Dalitz plot separation of the $I=1$ and $I=0$
parts of the $K\bar K \pi$ cross sections~\cite{BaBar:2007ceh}. The
BaBar $I=1$ cross sections provide a determination of the $I=1$ part of
the $K\bar{K}\pi$ contribution to $\rho_{EM}(s)$, and hence of the
$K\bar{K}\pi$ mode contributions to the lqc window quantities. The
corresponding $I=0$ contributions are obtained by subtracting the
BaBar-based $I=1$ results from the KNT19 $I=1+0$ totals. As an
example, we find, for the $K\bar{K}\pi$ contribution to the lqc
component of the LD window, the result
\beq
[a_\mu^{\rm LD}]^{\rm lqc}_{K\bar K \pi} = \frac{10}{9}
[a_\mu^{{\rm LD}}]^{I=1}_{K\bar K \pi} = 0.083(14)\times 10^{-10}.
\eeq
The corresponding results for the other windows are given in
Tab.~\ref{tab:all-amb-lqc}. Similarly, for the $K\bar{K}\pi$
contribution to the s+lqd component of the LD window, we find,
using the second line of Eq.~(\ref{eq:rhosplqd}),
\begin{align}
[a_\mu^{\rm LD}]^{\rm s+lqd}_{K\bar K \pi} &=
[a_\mu^{\rm LD}]_{K\bar K \pi}-\frac{10}{9}
[a_\mu^{{\rm LD}}]^{I=1}_{K\bar K \pi} \nonumber \\
&= [0.263(11) -\frac{10}{9}\times 0.075(13)]\units = 0.180(18) \units,
\end{align}
with the results for the other windows given in Tab.~\ref{tab:all-amb-splqd}.

For the $K\bar K2\pi$ channel, a modest in-principle improvement can
be achieved over the purely maximally conservative separation treatment by
first using BaBar's measurement of the $e^+e^- \to \phi \pi\pi$ cross
sections~\cite{BaBar:2011btv} and the PDG $\phi\rightarrow K\bar{K}$
branching fraction to quantify the purely
$I=0$ $e^+e^-\rightarrow \phi (\rightarrow K\bar{K})\pi\pi$ contribution
and then applying the maximally conservative separation treatment to the
rest of the $K\bar{K} 2\pi$ contributions.
In practice, the improvement is too small to make a significant
impact and the final uncertainties on the $K\bar{K}2\pi$ contributions
are still of the order of 100\%, as can be seen in Tabs.~\ref{tab:all-amb-lqc}
and~\ref{tab:all-amb-splqd}.

Finally, for the radiative modes $\pi^0\gamma$ and $\eta\gamma$, a
full decomposition into pure $I=1$, pure $I=0$ and MI components turns
out to be possible owing to the strong dominance of the observed
exclusive-mode-region cross sections by intermediate vector meson
contributions. We briefly outline the decomposition procedure below,
referring the reader to App.~B of Ref.~\cite{Benton:2023fcv} for further
details. We note first that the measured $e^+ e^-\rightarrow \pi^0\gamma$
and $e^+ e^-\rightarrow \eta\gamma$ cross sections display prominent narrow
$\omega$ and $\phi$ resonance peaks. The normalizations of the underlying
$V=\omega$ and $\phi$ contributions to the amplitudes are, of course,
set by the measured $V\rightarrow e^+ e^-$ and $V\rightarrow P\gamma$
($P=\pi ,\, \eta$) widths. Less immediately evident visually, but
necessarily also present, are broad $\rho$ contributions, with
normalizations set by the measured $\rho\rightarrow e^+ e^-$ and
$\rho\rightarrow P\gamma$ widths. In Ref.~\cite{Benton:2023fcv} it
was shown that, using PDG input for the above widths, the VMD
representations of the $e^+ e^-\rightarrow P\gamma$ amplitudes
produced by summing over the resulting externally determined
$V=\rho ,\, \omega$ and $\phi$ contributions accurately reproduce
the observed cross sections and, not surprisingly therefore, provide
accurate representation of the resulting
full HVP and
intermediate window integrals. Neglecting additional IB effects in the
photon-vector-meson couplings (since the $e^+ e^-\rightarrow \pi^0\gamma$
and $e^+ e^-\rightarrow \eta\gamma$ cross sections and associated
weighted integrals are already $O(\alpha_{EM})$ and hence first
order in IB), the $\rho$ contributions to the two amplitudes
come solely from the coupling of the $\rho$ to the $I=1$ (flavor $3$)
part of the EM current, and the $\omega$ and $\phi$ contributions
solely from the couplings of the $\omega$ and $\phi$ to the $I=0$
(flavor $8$) part. Thus, to first order in IB, the pure $I=1$
(flavor $33$) parts of the cross sections come from the squared
modulus of the $\rho$ contribution to the amplitude in question,
the pure $I=0$ (flavor $88$) part from the squared modulus of the
sum of the $\omega$ and $\phi$ contributions, and the MI (flavor
$38$) part from the interference between the $\rho$ and
$\omega +\phi$ contributions. The $\pi^0\gamma$ and $\eta\gamma$
contributions to the lqc and s+lqd components of the SD and
LD windows produced by the resulting $I=1$/$I=0$/MI
decompositions are listed in Tabs.~\ref{tab:all-amb-lqc}
and~\ref{tab:all-amb-splqd}, where, for completeness, we also
list the corresponding HVP and intermediate window results,
reported previously in Ref.~\cite{Benton:2023fcv}.

The small contributions from the remaining ambiguous exclusive
modes of the KNT19 data compilation,
are handled using the maximally conservative separation treatment,
and the results, again, collected in Tabs.~\ref{tab:all-amb-lqc}
and~\ref{tab:all-amb-splqd}.

\begin{table}[!t]
\begin{center}
\caption{\label{tab:all-amb-splqd}
Contributions of modes with no definite isospin to the s+lqd parts
of $a_\mu^{\rm SD}$, $a_\mu^{\rm int}$, $a_\mu^{\rm LD}$, and
$a_\mu^{\rm HVP}$. See text for the details of the treatment of
each channel. The label ``npp'' stands for ``non-purely pionic.''
Uncertainties in the last column take into account the
correlations between the three windows. All entries in units of $10^{-10}$. }
{\footnotesize
\begin{tabular}{lllll}
\toprule
mode &$[a_\mu^{\rm SD}]_X^{\rm s+lqd}\times 10^{10}$
& $[a_\mu^{\rm int}]_X^{\rm s+lqd}\times 10^{10}$
& $[a_\mu^{\rm LD}]_X^{\rm s+lqd}\times 10^{10}$
&$[a_\mu^{\rm tot}]_X^{\rm s+lqd}\times 10^{10}$ \\
\hline
$K\bar K$ & 3.229(39) & 18.55(17)
& 13.35(11) & 35.13(31) \\
$K\bar K \pi$ & 0.518(48) & 1.19(11)
& 0.180(18) & 1.89(18) \\
$\pi^0\gamma +\eta\gamma$ & 0.1763(66) & 1.50(6)
& 2.32(10) & 4.00(17) \\
$K\bar K 2\pi$ & 0.30(32) & 0.58(60)
& 0.062(62) & 0.94(98) \\
$K\bar K 3\pi$ & 0.0065(82) & 0.009(12)
& 0.00043(54) & 0.016(20) \\
low-s $\pi^0\gamma+\eta\gamma$ & 0.00043(54) &0.0066(82)
& 0.049(61) & 0.056(70) \\
$N\bar N$ & 0.010(12) & 0.016(19)
& 0.0012(15) & 0.026(33) \\
$\eta (\to{\rm npp}) K\bar K ({{\rm no}\, \phi})$ & 0.0021(27) & 0.0040(50)
& 0.00042(52) & 0.0065(82) \\
$\omega (\to{\rm npp})$ $K\bar K$ & 0.00068(85)
& 0.0010(12) & 0.000045(56) & 0.0017(21) \\
\hline
Total (s+lqd) & 4.25(33) & 21.86(64)
& 15.96(18) & 42.1(1.1) \\
\bottomrule
\end{tabular}}
\end{center}
\end{table}

\subsection{Perturbative contribution above
\boldmath \texorpdfstring{$s=(1.937~{\rm GeV})^2$}{s=(1.937 GeV)^2}}
\label{sec:pert}

In the inclusive region, i.e., for hadronic squared invariant masses
$s>(1.937~{\rm GeV})^2$, we use massless three-flavor pQCD
(we have checked that strange-quark mass corrections can safely be
neglected~\cite{Boito:2022rkw}). The Adler function is exactly known up to
$\mathcal{O}(\alpha_s^4)$~\cite{Baikov:2008jh} and we supplement it with an
estimate for the $\mathcal{O}(\alpha_s^5)$ coefficient, as described in our
previous works~\cite{Boito:2022rkw,Boito:2022dry,Benton:2023fcv}. To this
perturbative result, we add an estimate of the duality violating (DV)
contribution, which, essentially, captures the residual oscillations in the
spectral function due the tails of the higher-mass resonances. To
parametrize the DVs, we employ results from our previous study of the $I=1$
spectral function in $\tau\to {\rm hadrons}+\nu_\tau$~\cite{Boito:2020xli}
decays as well as knowledge about the $I=0$ contribution from
$e^+e^-\to {\rm hadrons}$~\cite{Boito:2018yvl}.

Perturbative QCD
is in good agreement with inclusive $R(s)$ data from
BES~\cite{BES:2001ckj,BES:2009ejh} and KEDR~\cite{KEDR:2018hhr}
for $s\geq 4~{\rm GeV}^2$, but in some tension with recent, more precise
results from BES-III~\cite{BESIII:2021wib} below charm threshold. Because
of this tension, and since the DV contribution represents an essential
limitation of perturbation theory, we have significantly enlarged
the error associated with the use of pQCD in the inclusive
region --- which is typically assessed using estimates of the size
of missing higher-orders in the $\alpha_s$ expansion. We consider,
therefore, as our final error on the inclusive-region
contribution the central value of the DV component. This procedure
produces a significantly increased error on the inclusive-region
contributions to the lqc and s+lqd components. For the lqc case,
the error is enlarged by factors that vary between 3 and 11 (depending
on the window) while for the s+lqd component the enlargement can be
up to factors of order 30. As we show here, because of the small
contribution from the inclusive region in essentially all cases (the
exception is the SD window, as could be expected) this increase in
error in the perturbative contribution has no meaningful impact
on the precision of our final results.

Since the details of our perturbative description in the inclusive
region, as well as of the parametrization of the DVs that we employ,
have already been extensively discussed in
Ref.~\cite{Boito:2022rkw,Benton:2023fcv} we, here, simply quote
the final numbers for the pQCD+DV contributions in
Tabs.~\ref{tab:final-results-lqc} and~\ref{tab:final-results-slqd}.

\subsection{Isospin-breaking corrections}
\label{sec:IB}

The final step is to estimate electromagnetic (EM) and
strong-isospin-breaking (SIB) contributions to the results discussed
above. These contributions must be subtracted before comparing
our results with those of isospin-symmetric lattice QCD. We follow the
treatment discussed extensively in our previous
works~\cite{Boito:2022rkw,Boito:2022dry,Benton:2023fcv} and here
simply summarize the application of this framework to the SD and
LD windows. Our isospin-symmetric results, as is the case for many
lattice groups, correspond to a definition of the isospin limit
of QCD in which all pions have the neutral pion mass.

We work to first order in IB and start from the observation that, to this
precision, SIB appears only in the MI component of $\rho_{\rm EM}(s)$
while EM contributions appear in all of the pure $I=1$, pure $I=0$
and MI components. MI contributions, in general, produce small
IB ``contaminations'' of the nominally pure $I=1$ $G$-parity positive
and nominally pure $I=0$ $G$-parity negative exclusive-mode contributions
discussed above. These must be subtracted, mode-by-mode, to arrive at
data-driven isospin-limit lqc and s+lqd results suitable for comparison
to the corresponding lattice results. It is expected, however,
that the dominant such contaminations will be those in the
$2\pi$ and $3\pi$ channels, which are strongly enhanced by the
effects of $\rho-\omega$ mixing through the processes
$e^+e^-\to\omega\to\rho\to 2\pi$ and $e^+e^-\to\rho\to\omega\to 3\pi$.
To estimate the resulting dominant MI contaminations
we use the results of
Ref.~\cite{Hoferichter:2023sli,Colangelo:2022prz,Hoferichter:2023bjm}
where the $2\pi$ and $3\pi$ electroproduction data were fitted with
dispersive representations incorporating the effects of
$\rho-\omega$ mixing. Since these estimates are obtained using
experimental input, they, of course, include both the EM and SIB
components of the MI contributions.

The resulting estimates for the MI components of the nominally
$I=1$ $2\pi$-mode contributions to HVP integral and the three windows
discussed in this paper can be found in Tab.~I of
Ref.~\cite{Hoferichter:2023sli} and read
\begin{align}
&[a_\mu^{\rm SD}]_{\pi\pi}^{\rm MI}\times 10^{10} = 0.06(1),\nn\\
&[a_\mu^{\rm int}]_{\pi\pi}^{\rm MI}\times 10^{10} = 0.86(6), \nn\\
&[a_\mu^{\rm LD}]_{\pi\pi}^{\rm MI}\times 10^{10} = 2.87(12),\nn\\
&[a_\mu^{\rm HVP}]_{\pi\pi}^{\rm MI}\times 10^{10} = 3.79(19).
\label{eq:2pi-MI}
\end{align}
In the $2\pi$ channel, this MI contribution, in spite of the
enhancement due to the $\rho-\omega$ mixing, never exceeds 0.85\%
of the $2\pi$ total given in Tab.~\ref{tab:unambiguous-modes-I1}.
With no analogous narrow, nearby resonance enhancements of this
type expected for other nominally $I=1$ modes, we consider it
safe to assume that the total MI contamination present in
the contributions from these other modes will not exceed $1\%$ of
the sum of their contributions. In this spirit, we add to the results of
Eq.~(\ref{eq:2pi-MI}) an additional uncertainty equal to 1\%
of the sum of all non-2$\pi$ nominally $I=1$ contributions, from
both unambiguous and ambiguous modes (the latter with the exception
of the radiative modes, $\pi^0\gamma$ and $\eta\gamma$, where
the VMD representation provides a full $I=1$/$I=0$/MI separation and there
is, thus, no MI contamination of either the $I=1$ or $I=0$ contribution).
The non-$2\pi$, $I=1$ totals for each window, obtained by adding to
the unambiguous-mode $I=1$ results of Tab.~\ref{tab:unambiguous-modes-I1}
the $I=1$ components of the ambiguous-mode contributions that can be
inferred from Tab.~\ref{tab:all-amb-lqc} (excluding the
radiative channels $\pi^0\gamma$ and $\eta\gamma$), are the following
\begin{align}
&[a_\mu^{\rm SD}]_{{\rm non}-2\pi}^{I=1}\times 10^{10} = 9.12(34),\nn\\
&[a_\mu^{\rm int}]_{{\rm non}-2\pi}^{I=1}\times 10^{10} = 25.66(76), \nn\\
&[a_\mu^{\rm LD}]_{{\rm non}-2\pi}^{I=1}\times 10^{10}= 6.68(17),\nn\\
&[a_\mu^{\rm tot}]_{{\rm non}-2\pi}^{I=1}\times 10^{10} = 41.5(1.2).
\label{eq:non-2pi-totals}
\end{align}
Our final estimates for the nominally $I=1$ MI contaminations
consist of the numbers given in Eq.~(\ref{eq:2pi-MI}) with 1\% of the
central values of Eq.~(\ref{eq:non-2pi-totals}) added in quadrature
as an additional uncertainty. These results are to multiplied by
$10/9$ to convert them to the corresponding lqc MI contaminations and
subtracted from the sum of the uncorrected lqc results obtained above,
shown in the first three lines of Tab.~\ref{tab:final-results-lqc}.
The resulting MI corrections, to be added to the other entries, are
listed in line four of this table.

The procedure employed to estimate the $I=0$ MI IB contribution is
very similar: we use the results of~\rcite{Hoferichter:2023sli} as
estimates of the MI contaminations, $[a_\mu^{\rm win}]_{3\pi}^{\rm MI}$,
present in the nominally $I=0$ $3\pi$ window contributions and add to
those results an additional uncertainty of 1\% of the total of
all non-3$\pi$, nominally $I=0$ contributions, both unambiguous and
ambiguous --- again with the exception of those from the radiative
modes $\pi^0\gamma$ and $\eta\gamma$, where the MI contributions has
been explicitly determined and the $I=0$ contributions determined
using the VMD representation contain no MI contamination. The $3\pi$
MI IB contributions from Tab.~I of Ref.~\cite{Hoferichter:2023sli} are
\begin{align}
&[a_\mu^{\rm SD}]_{3\pi}^{\rm MI}\times 10^{10} = -0.13(3),\nn\\
&[a_\mu^{\rm int}]_{3\pi}^{\rm MI}\times 10^{10} =-1.03(27) , \nn\\
&[a_\mu^{\rm LD}]_{3\pi}^{\rm MI}\times 10^{10} =-1.52(40) ,\nn\\
&[a_\mu^{\rm HVP}]_{3\pi}^{\rm MI}\times 10^{10} =-2.68(70) .
\label{eq:3pi-MI}
\end{align}
The corresponding $I=0$ non-$3\pi$ totals, obtained by adding to the
results of Tab.~\ref{tab:unambiguous-modes-I0} the $I=0$ ambiguous-mode
contributions inferable from Tabs.~\ref{tab:all-amb-splqd} and
\ref{tab:all-amb-lqc} (again excluding the $\pi^0\gamma$ and
$\eta\gamma$ mode contributions) are found to be
\begin{align}
&[a_\mu^{\rm SD}]_{{\rm non}-3\pi}^{I=0}\times 10^{10} = 5.04(30) ,\nn\\
&[a_\mu^{\rm int}]_{{\rm non}-3\pi}^{I=0}\times 10^{10} =22.53(59) , \nn\\
&[a_\mu^{\rm LD}]_{{\rm non}-3\pi}^{I=0}\times 10^{10}= 14.00(14) ,\nn\\
&[a_\mu^{\rm HVP}]_{{\rm non}-3\pi}^{I=0}\times 10^{10}=41.56(97).
\label{eq:non-3pi-totals}
\end{align}
Adding 1\% of these non-3$\pi$, $I=0$ totals as an additional
uncertainty to the results of Eq.~(\ref{eq:3pi-MI}) and combining
these results with the $I=1$ MI IB contributions as per
Eq.~(\ref{eq:rhosplqd}), we find the s+lqd MI corrections shown
in the 5-th row of Tab.~\ref{tab:final-results-slqd}.

With the full EM+SIB MI corrections in hand, the remaining IB
effects to be dealt with are the EM corrections to the pure $I=1$ and
$I=0$ window contributions. At this point, although several EM
contributions have been estimated from experimental
data~\cite{Hoferichter:2023bjm,Colangelo:2022prz,Hoferichter:2023sli},
other potentially non-negligible EM effects have not (see the discussion
in the appendix of \rcite{Boito:2022dry}). We have decided,
therefore, to rely on lattice EM data for our estimates of the $I=1$
and $I=0$ EM corrections. As discussed below, these corrections
are very small in the lqc case and completely negligible for
the s+lqd components, which means that our final results are still
(almost) purely data-driven.

We start with a discussion of the lqc EM contributions. For the total
HVP, the EM contribution was published in 2021 by
BMW~\cite{Borsanyi:2020mff} and corrected in their more recent
paper~\cite{Boccaletti:2024guq}. The corrected, 2024 result,
\beq
\Delta_{\rm EM}a_\mu^{\rm HVP, lqc}\times 10^{10}
= -1.57(42)(35)=-1.57(55),
\eeq
has to be subtracted from our HVP result before
comparison with the isospin-symmetric lattice HVP value.
BMW also provided the EM correction to the intermediate window
lqc component~\cite{Borsanyi:2020mff},
\beq
\Delta_{\rm EM}a_\mu^{\rm int, lqc}\times 10^{10} = -0.035(59),
\eeq
which constitutes a very small correction with essentially no impact
on the final results of our previous
works~\cite{Benton:2023dci,Benton:2023fcv}.

For the SD and LD windows no number was provided in the BMW papers. This
leaves a contribution of about $1.535\units$ to be split between these
two windows. Here, we make the assumption that the EM contribution to the
lqc component of the SD window is negligible, which means that the
entirety of the remaining $1.535\units$ is attributed to the LD window.
Two arguments support this assumption. First, a significant portion
of the EM contribution to the SD window is amenable to a perturbative
calculation. This gives only a tiny correction, which the Mainz
collaboration estimated to be 0.03$\units$~\cite{Kuberski:2024bcj}. Second,
the same Mainz paper quotes an initial direct lattice simulation result for
the SD EM contribution equal to $0.15(15)\%$ of the sum of the corresponding
light and strange connected contributions, corresponding to a central
value of $0.085\units$~\cite{Kuberski:2024bcj}. This is compatible
(though smaller in magnitude) with what we expect to be a conservative
estimated bound on the magnitude of the total EM contribution, equal
to $\alpha_{\rm EM}$ times the total of the unambiguous and ambiguous
exclusive-mode SD lqc contributions plus $\alpha_{\rm EM}/\pi$ times
the corresponding pQCD contribution. This yields the bound
\begin{align}
\vert \Delta_{\rm EM} a_\mu^{\rm SD,lqc} \vert&
\leq
\alpha_{EM} \left(
[a_\mu^{\rm SD}]_{\rm unamb}^{\rm lqc}
+[a_\mu^{\rm SD}]_{\rm amb}^{\rm lqc}\right)
+\frac{\alpha_{\rm EM}}{\pi}[a_\mu^{\rm SD}]_{\rm pQCD}^{\rm lqc}\nn \\
&=\alpha_{\rm EM}(25.99+0.75)\times 10^{-10} +
\frac{\alpha_{\rm EM}}{\pi}20.28\times 10^{-10} \nn \\
&= 0.24\times 10^{-10},\label{eq:EMIBSD}
\end{align}
where $[a_\mu^{\rm SD,lqc}]_{\rm unamb}$,
$[a_\mu^{\rm SD,lqc}]_{\rm amb}$, and $[a_\mu^{\rm SD}]_{\rm pQCD}^{\rm lqc}$
are, respectively, the unambiguous-mode, ambiguous-mode and the
pQCD contributions to $a_\mu^{\rm SD,lqc}$, given in
Tab.~\ref{tab:final-results-lqc}. This result is a factor of nearly
$3$ times larger than the uncertainty on the Mainz result. To be
conservative, we assign this larger estimate as the uncertainty
on our EM SD assumption.
Our final estimates for the EM IB contribution to the lqc SD and LD windows
are then
\begin{align}
&\Delta_{\rm EM}a_\mu^{\rm SD,lqc}\times 10^{10} = 0.00(24), \nn\\
&\Delta_{\rm EM}a_\mu^{\rm LD,lqc}\times 10^{10} = 1.54(55) .
\end{align}

For the s+lqd component, the EM corrections to the intermediate window
and to the total HVP can be obtained from the published BMW results
via the diagrammatic approach explained in detail in
Ref.~\cite{Boito:2022rkw}. These corrections turn out to be tiny due, in
part, to strong cancellations in the numerically dominant contributions
arising from light-quark EM valence-valence connected and disconnected
diagrams. For the intermediate window, for example, we obtain the
following EM correction to the s+lqd component~\cite{Benton:2023dci}
\begin{align}
&\Delta_{\rm EM}a_\mu^{\rm int,s+lqd}\times 10^{10} = 0.012(11) ,
\end{align}
more than 60 times smaller than our final error. Since the final relative
uncertainties in the s+lqd SD and LD windows are larger than that of
the intermediate window, and since the EM contributions to s+lqd
intermediate window and total HVP results are so small, we believe
it to be safe to neglect the EM corrections for the s+lqd component
in the case of the SD and LD windows as well.

\section{Final results}
\label{sec:final-results}

We are now in a position to obtain our final results based on the
pre-CMD-3 KNT-19 data compilation. Tabs.~\ref{tab:final-results-lqc}
and~\ref{tab:final-results-slqd} collect all the partial contributions
discussed in the previous sections and give, in their last rows, the
final results for the isospin-symmetric lqc and s+lqd components of
the SD, the intermediate, and the LD windows, as well as of the total
HVP. The results for the SD and LD windows are new, while results for
the intermediate window and for the total HVP, apart from small updates
having very little numerical impact, were already discussed in
Refs.~\cite{Boito:2022dry,Boito:2022rkw,Benton:2023dci,Benton:2023fcv}.

\begin{table}[!t]
\begin{center}
\caption{\label{tab:final-results-lqc}
KNT19-based~\cite{Keshavarzi:2019abf} results for the lqc component of
$a_\mu^{\rm SD}$, $a_\mu^{\rm int}$, $a_\mu^{\rm LD}$, and $a_\mu^{\rm HVP}$.
The last row gives the final isospin-symmetric results to be compared
with lattice-QCD. Uncertainties in the last column take into account
the correlations between the three windows. All entries in units of
$10^{-10}$.}
\begin{tabular}{lllll}
\toprule
&$a_\mu^{\rm SD,lqc}\times 10^{10}$ & $a_\mu^{\rm int,lqc}\times 10^{10}$
& $a_\mu^{\rm LD,lqc}\times 10^{10}$ &$a_\mu^{\rm HVP,lqc}\times 10^{10}$ \\
\hline
unamb. modes & 25.99(21) & 186.94(80)
& 390.6(1.6) & 603.6(2.3) \\
amb. modes & 0.75(32) & 1.88(61)
& 0.613(91) & 3.2(1.0) \\
pt. QCD + DVs & 20.28(0.10) & 11.06(0.16)
& 0.346(11) & 31.68(28) \\
MI IB correction & $-0.07(10)$
& $-0.96(0.30)$& $-3.19(15)$ & $-4.21(51)$ \\
EM IB correction& 0.00(24) & 0.035(59)
& 1.54(55) & 1.57(55) \\
\hline
Total & 46.96(48)
& 199.0(1.1) & 389.9(1.7) & 635.8(2.6) \\
\bottomrule
\end{tabular}
\end{center}
\end{table}

\begin{figure}[!t]
\begin{center}
\includegraphics[width=0.49\textwidth]{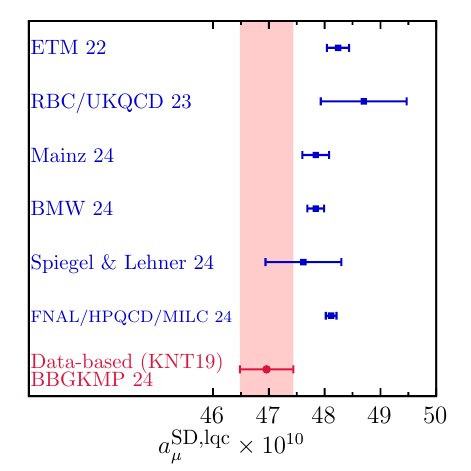}
\includegraphics[width=0.49\textwidth]{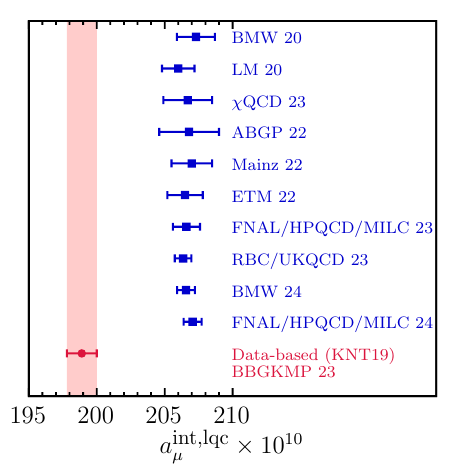}
\includegraphics[width=0.49\textwidth]{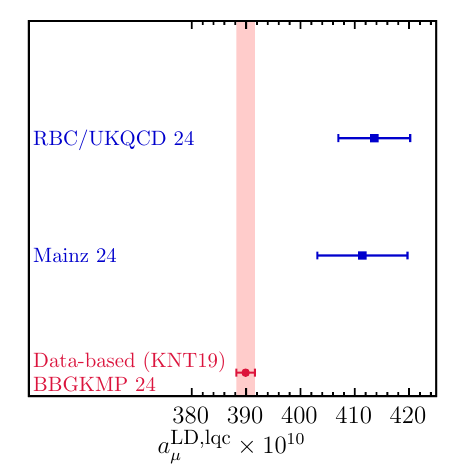}
\includegraphics[width=0.49\textwidth]{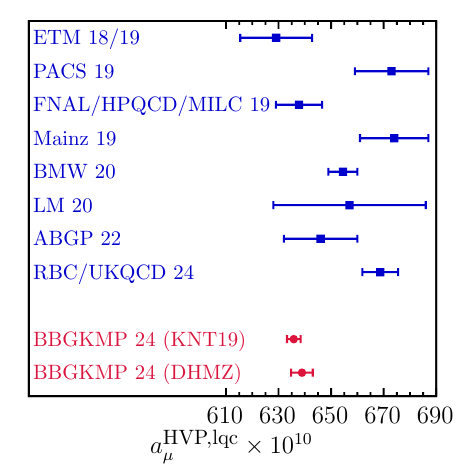}
\caption{Data-driven KNT19-based~\cite{Keshavarzi:2019abf} results for
the lqc components of the RBC/UKQCD windows compared with isospin-symmetric
lattice-QCD determinations of the same quantities. (Upper-left panel)
Data-driven $a_\mu^{\rm SD,lqc}$ compared with results from
Refs.~\cite{ExtendedTwistedMass:2022jpw,RBC:2023pvn,Kuberski:2024bcj,Boccaletti:2024guq,Spiegel:2024dec,Bazavov:2024xsk}. Results from this work are labeled ``BBGKMP 24.''
(Upper-right panel) Data-driven $a_\mu^{\rm int,lqc}$ compared with results
from Refs.~\cite{Borsanyi:2020mff,Lehner:2020crt,Wang:2022lkq,Aubin:2022hgm,Ce:2022kxy,ExtendedTwistedMass:2022jpw,FermilabLatticeHPQCD:2023jof,RBC:2023pvn,Bazavov:2024xsk}.
(Bottom-left panel) Data-driven $a_\mu^{\rm LD,lqc}$ compared with the recent
results of Refs.~\cite{RBC:2024fic,Djukanovic:2024cmq}. (Bottom-right panel) Data-driven $a_\mu^{\rm HVP,lqc}$ compared with results from Refs.~\cite{Aubin:2022hgm,Borsanyi:2020mff,Lehner:2020crt,Gerardin:2019rua,FermilabLattice:2019ugu,Giusti:2018mdh,Giusti:2018vrc,Shintani:2019wai,RBC:2024fic}.}
\label{fig:results-lqc-KNT19}
\end{center}
\end{figure}

In the lqc results of Tab.~\ref{tab:final-results-lqc} for
the intermediate and LD windows, as well as for the total HVP, the
unambigous-mode contribution overwhelmingly dominates, with an important
fraction of the final results arising from the 2$\pi$ channel
contribution, which represents 80\% for the intermediate
window and total HVP results and 88\% for the LD window. The lqc
components of these windows are, therefore, very sensitive to any issue
related to the data-driven contribution from $e^+e^-\to \pi^+\pi^-$.
In the SD window, although the contribution from unambiguous modes is
still dominant, the pQCD contribution is large (amounting
to 43\% of the final result), while pQCD gives only a small
contributions in all other cases. In Fig.~\ref{fig:results-lqc-KNT19}
we compare our final results for the lqc component of the three
RBC/UKQCD windows to isospin-symmetric lattice-QCD results.
In these results, we see that for $a_\mu^{\rm int,lqc}$, as already
discussed in Refs.~\cite{Benton:2023dci,Benton:2023fcv}, there is a
greater-than-5$\sigma$ discrepancy with the most precise lattice
results, including the new BMW result~\cite{Boccaletti:2024guq}. The
recently unblinded results from RBC/UKQCD~\cite{RBC:2024fic} and Mainz~\cite{Djukanovic:2024cmq} for the lqc component of the LD
window (in the ``BMW world'') also show 3.5$\sigma$ and 2.5$\sigma$ discrepancies, respectively,
with respect to our KNT19-based data-driven result. In the SD window,
our data-based lqc component, while systematically lower,
differs less significantly from
the lattice determinations.
We recall that the SD window receives an important pQCD contribution,
and, since we use perturbation theory, the final result in
this case is less dominated by exclusive-mode experimental data.
Given the reduced role of $\rho$-region $2\pi$ exclusive-mode
contributions for the SD window, the reduced difference between the
data-driven and lattice results in the SD case is also compatible
with the hypothesis that all of the observed discrepancies between
data-driven and lattice results have their source in contributions
from the $\rho$ peak region.

\begin{table}[!t]
\begin{center}
\caption{\label{tab:final-results-slqd}
KNT19-based~\cite{Keshavarzi:2019abf} results for the s+lqd component of
$a_\mu^{\rm SD}$, $a_\mu^{\rm int}$, $a_\mu^{\rm LD}$, and $a_\mu^{\rm HVP}$.
The last row gives the final isospin-symmetric results to be compared
with lattice-QCD. Uncertainties in the last column take into account the
correlations between the three windows. All entries in units of $10^{-10}$. }
{\small
\begin{tabular}{lllll}
\toprule
&$a_\mu^{\rm SD,s+lqd}\times 10^{10}$ & $a_\mu^{\rm int,s+lqd}\times 10^{10}$
& $a_\mu^{\rm LD,s+lqd}\times 10^{10}$ &$a_\mu^{\rm HVP,s+lqd}\times 10^{10}$\\
\hline
unamb. total & 0.903(68) & 1.99(38)
&$-13.32(57)$ & $-10.43(98)$\\
amb. total & 4.25(33) & 21.86(64)
& 15.96(18) & 42.1(1.1) \\
pt. QCD + DVs & 3.93(11) & 2.00(17)
& 0.054(13) & 5.99(30) \\
MI IB correction & 0.137(60) & 1.13(35)
& 1.84(42) & 3.10(82) \\
\hline
Total & 9.21(36)
& 26.98(84) & 4.53(73)& 40.7(1.7) \\
\bottomrule
\end{tabular}}
\end{center}
\end{table}

We turn now to the results for the s+lqd components,
given in Tab.~\ref{tab:final-results-slqd}.
There are three recent lattice determinations of the relevant components of the SD window~\cite{Kuberski:2024bcj,Boccaletti:2024guq,Bazavov:2024xsk}, all in good agreement. BMW~\cite{Boccaletti:2024guq}
finds, for the strange-quark-connected and light-quark-disconnected
contributions, the results, $9.04(7)\times 10^{-10}$ and
$-0.0007(102)\times 10^{-10}$, respectively. The corresponding correlation
is not quoted, so we know only that the uncertainty on the
BMW s+lqd sum must lie in the range $(0.07\pm 0.01)\times 10^{-10}$.
The Mainz collaboration~\cite{Kuberski:2024bcj} finds a compatible result,
$9.072(59)\times 10^{-10}$, for the strange-quark connected contribution,
and comments that the strange-quark- and light-quark-disconnected contributions were found
to be irrelevant at the level of precision of the strange connected result. Finally, FNAL/HPQCD/MILC quotes results $9.103(3)(21)\times 10^{-10}$
and $-0.0002(6)(53)\times 10^{-10}$ for the strange-quark-connected and
light-quark-disconnected contributions, with a $0.13$ correlation coefficient between them~\cite{Bazavov:2024xsk}.

Taking the most conservative assessment of the BMW error, and
following Mainz in neglecting their disconnected contribution, we find for
the lattice versions of the s+lqd component of the SD window, the results
\begin{align}
a_\mu^{\rm SD,s+lqd}\times 10^{10} &= 9.04(8) \qquad\,\,\,\,\,\, (\mbox{BMW 24~\cite{Boccaletti:2024guq}}),\nonumber \\
a_\mu^{\rm SD,s+lqd}\times 10^{10} &= 9.072(59) \qquad (\mbox{Mainz 24~\cite{Kuberski:2024bcj}}), \nonumber \\
a_\mu^{\rm SD,s+lqd}\times 10^{10} &= 9.103(22)\qquad (\mbox{FNAL/\-HPQCD/\-MILC 24~\cite{Bazavov:2024xsk}}).
\end{align}
These are all in excellent agreement with our data-driven determination,
$a_\mu^{\rm SD, s+lqd}\, \times 10^{-10} = 9.21(36)$, though in this
case the data-driven uncertainty is significantly larger than that on
the lattice results. A visual account of this
comparison is given in the left-hand panel of Fig.~\ref{fig:results-splqd-KNT19}.
For the s+lqd contribution to the LD window, we obtain
$4.53(73)\units$. The only lattice determination of this quantity, by the Mainz collaboration, gives
\begin{equation}\label{eq:Mainz-LD-splqd}
a_\mu^{\rm LD,s+lqd}\times 10^{10}=1.3(2.4)\qquad(\mbox{Mainz 24~\cite{Kuberski-private-comm,Djukanovic:2024cmq}}),
\end{equation}
which is compatible, within $1.3\sigma$, with our data-driven result.
The results for $a_\mu^{\rm int,s+lqd}$ and $a_\mu^{\rm HVP,s+lqd}$, in turn, agree well with the lattice determinations of
Refs.~\cite{Borsanyi:2020mff,Budapest-Marseille-Wuppertal:2017okr,RBC:2018dos,Blum:2015you} and Refs.~\cite{RBC:2018dos,Borsanyi:2020mff,Ce:2022kxy,ExtendedTwistedMass:2022jpw}, respectively, as discussed in our previous works~\cite{Boito:2022rkw,Benton:2023fcv}. The right-hand panel of Fig.~\ref{fig:results-splqd-KNT19} shows an updated visual account of this comparison for $a_\mu^{\rm int,s+lqd}$ including the new result of Ref.~\cite{Bazavov:2024xsk}, also in very good agreement with our data-driven determination.

\begin{figure}[!t]
\begin{center}
\includegraphics[width=0.49\textwidth]{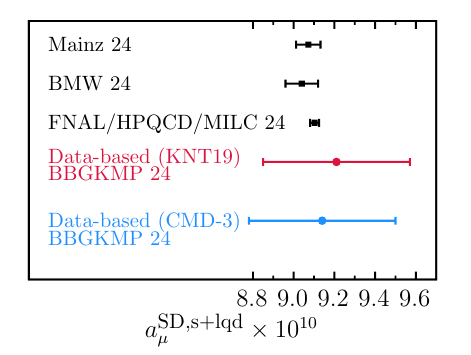}
\includegraphics[width=0.49\textwidth]{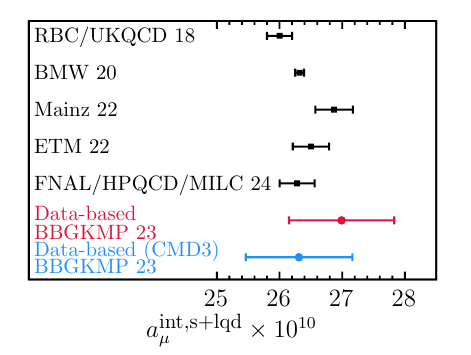}
\caption{Data-driven results for
the s+lqd components of the RBC/UKQCD SD and intermediate windows compared with isospin-symmetric lattice determinations of the same quantities. (Left-hand panel) Data-driven results for $a_\mu^{\rm SD, s+lqd}$ compared with results from Refs.~\cite{Bazavov:2024xsk,Kuberski:2024bcj,Boccaletti:2024guq}. See the text for a discussion of how the lattice numbers were obtained. (Right-hand panel) Data-driven results for $a_\mu^{\rm int, s+lqd}$ compared with results from Refs.~\cite{Bazavov:2024xsk,Borsanyi:2020mff,RBC:2018dos,Ce:2022kxy,ExtendedTwistedMass:2022jpw}.
Results from this work are labeled ``BBGKMP~24.''}
\label{fig:results-splqd-KNT19}
\end{center}
\end{figure}

We close this section with a short discussion of the correlations
between the results for the different windows.
Clearly, results for the same component (lqc or s+lqd) of the
different windows are expected to be highly correlated, since they
are simply differently weighted integrals of the same (lqc
or s+lqd) spectral data. Although the central values of the final
window results in Tabs.~\ref{tab:final-results-lqc}
and~\ref{tab:final-results-slqd} can be directly summed to
reproduce the $\amuHVP$ values in the last column, combining the
errors on the three window quantities to obtain that on $\amuHVP$
requires knowledge of the correlations between the window errors.

The
correlation between a pair of such window results is straightforwardly
obtained from the errors on the two individual windows (given in
Tabs.~\ref{tab:final-results-lqc} and~\ref{tab:final-results-slqd}
above) in combination with a separate, direct evaluation of the
error on the sum of the two windows. As input to this assessment,
we have assumed that the MI IB contributions obtained from
Ref.~\cite{Hoferichter:2023sli} are 100\% correlated, which is
both conservative and supported by the uncertainties given in the
original reference. The results, however, do not include
effects due to the $I=1$ and $I=0$ EM IB corrections, which do,
however, enter the computation of the lqc components. EM
contributions to the errors on the sums of window pairs, which
would be needed to incorporate these effects, cannot be reliably
included because the information available from
Refs.~\cite{Borsanyi:2020mff,Boccaletti:2024guq}, though including
the central values and errors for the intermediate window and total
HVP, does not include the corresponding correlation. We do not,
however, expect this limitation to be numerically relevant since
the contribution of the EM IB corrections to the final central
values and uncertainties are small.

With the strategy described above, we obtain the following
non-trivial correlation coefficients for the lqc components
of the three RBC/UKQCD windows
\begin{align}
\rho_{\rm SD,int} &= 0.894, \nn\\
\rho_{\rm int,LD} &= 0.509 , \nn\\
\rho_{\rm SD,LD} &= 0.341.
\label{eq:finalcorrelations-lqc}
\end{align}
The results for the correlations of the s+lqd components are
\begin{align}
\rho_{\rm SD,int}^{\rm s+lqd} &= 0.888, \nn\\
\rho_{\rm int,LD}^{\rm s+lqd}&= 0.687 , \nn\\
\rho_{\rm SD,LD}^{\rm s+lqd} &= 0.228.
\label{eq:finalcorrelations-slqd}
\end{align}
With these correlation coefficients and standard error propagation,
one can verify that the addition of the three window results does
reproduce the HVP totals shown in the fifth columns
of Tabs.~\ref{tab:final-results-lqc} and~\ref{tab:final-results-slqd}.

\section{Potential impact of CMD-3 results}
\label{sec:CMD-3-results}

\begin{table}[!t]
\begin{center}
\caption{\label{tab:final-results-lqc-CMD3} Exploratory results for the
lqc component of $a_\mu^{\rm SD}$, $a_\mu^{\rm int}$, $a_\mu^{\rm LD}$,
and $a_\mu^{\rm HVP}$ using the CMD-3 $\pi^+\pi^-$ data~\cite{CMD-3:2023alj}
in the energy region covered by the CMD-3 experiment and KNT19
data~\cite{Keshavarzi:2019abf} otherwise. Uncertainties
in the last column take into account the correlations between
the three windows. All entries in units of $10^{-10}$. }
\begin{tabular}{lllll}
\toprule
&$a_\mu^{\rm SD,lqc}\times 10^{10}$ & $a_\mu^{\rm int,lqc}\times 10^{10}$
& $a_\mu^{\rm LD,lqc}\times 10^{10}$ &$a_\mu^{\rm HVP,lqc}\times 10^{10}$ \\
\hline
unambiguous total & 26.67(25) & 193.8(1.5)
& 407.2(3.2) & 627.7(4.8) \\
ambiguous total & 0.75(32) & 1.88(61)
& 0.613(91) & 3.2(1.0) \\
pt. QCD + DVs & 20.28(0.10) & 11.06(0.16)
& 0.346(11) & 31.68(28) \\
MI IB correction & $-0.07(10)$
& $-0.96(0.30)$ & $-3.19(15)$ & $-4.21(51)$ \\
EM IB correction & 0.00(24) & 0.035(59)
& 1.54(55) & 1.57(55) \\
\hline
Total & 47.63(50)
& 205.8(1.6) & 406.5(3.2) & 660.0(4.9) \\
\bottomrule
\end{tabular}
\end{center}
\end{table}

\begin{table}[!h]
\begin{center}
\caption{\label{tab:final-results-slqd-CMD3}
Exploratory results for the s+lqd component of $a_\mu^{\rm SD}$,
$a_\mu^{\rm int}$, $a_\mu^{\rm LD}$, and $a_\mu^{\rm HVP}$ using the
CMD-3 $\pi^+\pi^-$ data~\cite{CMD-3:2023alj} in the energy region
covered by the CMD-3 experiment and KNT19
data~\cite{Keshavarzi:2019abf} otherwise.
Uncertainties in the last column take into account the correlations
between the three windows. All entries in units of $10^{-10}$. }
{\small
\begin{tabular}{lllll}
\toprule
&$a_\mu^{\rm SD,s+lqd}\times 10^{10}$ & $a_\mu^{\rm int,s+lqd}\times 10^{10}$
&$a_\mu^{\rm LD,s+lqd}\times 10^{10}$ &$a_\mu^{\rm HVP,s+lqd}\times 10^{10}$ \\
\hline
unamb. total & 0.836(69) & 1.31(40)
&$-14.89(63)$ & $-12.7(1.1)$\\
amb. total & 4.25(33) & 21.86(64)
& 15.96(18) & 42.1(1.1) \\
pt. QCD + DVs & 3.93(11) & 2.00(17)
& 0.054(13) & 5.99(30) \\
MI IB correction & 0.137(60) & 1.13(35)
& 1.84(42) & 3.10(82) \\
\hline
Total & 9.14(36)
& 26.30(85) & 2.96(79) & 38.4(1.7) \\
\bottomrule
\end{tabular}}
\end{center}
\end{table}

\begin{figure}[!t]
\begin{center}
\includegraphics[width=0.49\textwidth]{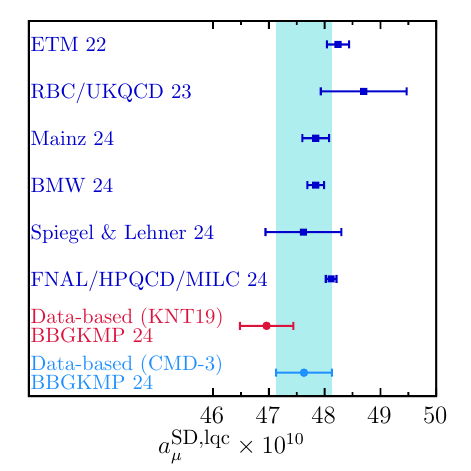}
\includegraphics[width=0.49\textwidth]{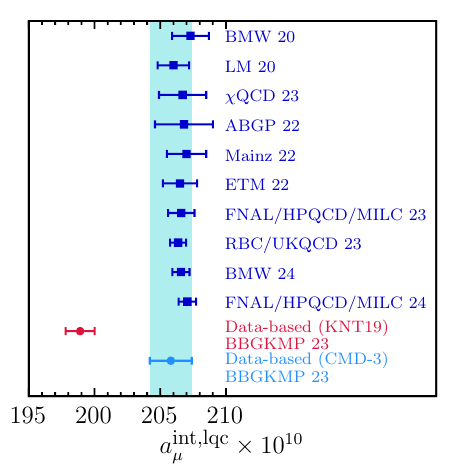}
\includegraphics[width=0.49\textwidth]{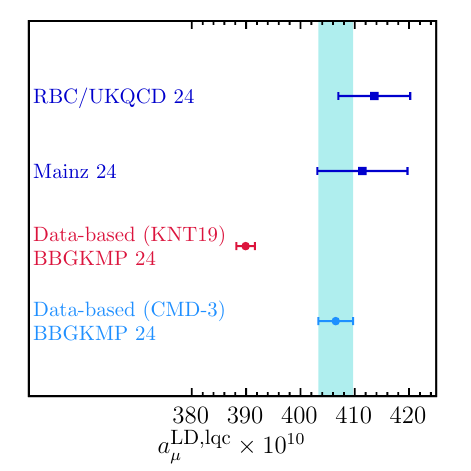}
\includegraphics[width=0.49\textwidth]{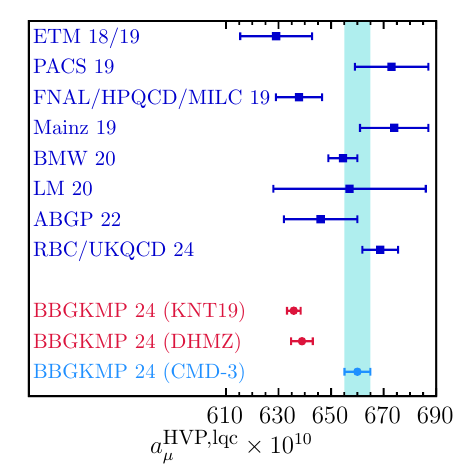}
\caption{Exploratory data-driven results for the lqc components
of the RBC/UKQCD windows employing CMD-3 $2\pi$ data~\cite{CMD-3:2023alj}
in the energy region covered by the CMD-3 experiment and KNT19
data~\cite{Keshavarzi:2019abf} elsewhere (light blue) in comparison
with data-driven results obtained employing KNT19 data only (red), and
isospin-symmetric lattice-QCD determinations of the same quantities. Results from this work are labeled ``BBGKMP~24.''
(Upper-left panel) Data-driven results for $a_\mu^{\rm SD,lqc}$
compared with results from
Refs.~\cite{ExtendedTwistedMass:2022jpw,RBC:2023pvn,Kuberski:2024bcj,Boccaletti:2024guq,Spiegel:2024dec,Bazavov:2024xsk}.
(Upper-right panel) Data-driven results for $a_\mu^{\rm int,lqc}$
compared with results from
Refs.~\cite{Borsanyi:2020mff,Lehner:2020crt,Wang:2022lkq,Aubin:2022hgm,Ce:2022kxy,ExtendedTwistedMass:2022jpw,FermilabLatticeHPQCD:2023jof,RBC:2023pvn,Bazavov:2024xsk}.
(Bottom-left panel) Data-driven results for $a_\mu^{\rm LD,lqc}$
compared with the recent results of Refs.~\cite{RBC:2024fic,Djukanovic:2024cmq}.
(Bottom-right panel) Data-driven $a_\mu^{\rm HVP,lqc}$ compared with results from Refs.~\cite{Aubin:2022hgm,Borsanyi:2020mff,Lehner:2020crt,Gerardin:2019rua,FermilabLattice:2019ugu,Giusti:2018mdh,Giusti:2018vrc,Shintani:2019wai,RBC:2024fic}.}
\label{fig:results-lqc-CMD3}
\end{center}
\end{figure}

In this section we investigate the potential impact of the new CMD-3
$e^+e^-\to \pi^+\pi^-$ cross-sections~\cite{CMD-3:2023alj} on the results
discussed in the previous section. The CMD-3 cross sections, as is
now well known, are significantly higher than those of earlier
experiments in the region around the $\rho$ peak, which gives a very
prominent contribution to $\amuHVP$. The shift in the $2\pi$
contribution produced by the CMD-3 data would, in fact, be
sufficient to eliminate the discrepancy between the experimental
result for $a_\mu$ and the White Paper SM expectation obtained
using the pre-CMD-3 dispersive HVP result. Unfortunately, the
disagreement between the $e^+e^-\to \pi^+\pi^-$ cross sections
obtained by CMD-3 and earlier experiments is sufficiently large that
a meaningful combination of all $2\pi$-channel data appears impossible,
at present. Given this state of affairs, we perform, in this section,
a preliminary exploration, in the spirit of our previous
work~\cite{Benton:2023fcv}, in which KNT19 $2\pi$ $R(s)$
data is simply replaced by CMD-3 results in the region covered by the
CMD-3 experiment, {\it i.e.}, for center-of-mass energies in the
range 0.327~GeV to 1.199~GeV. For this exercise, we apply vacuum
polarization (VP) corrections to the physical cross sections obtained
from the CMD-3 pion form factor results and dress the resulting
bare cross sections with final state radiation (FSR) correction
factors, using the same VP and FSR corrections factors employed by
CMD-3.\footnote{We thank Fedor Ignatov for providing the VP
corrections used by CMD-3.}

The results of this exploration
for the KNT19 exclusive-mode-region $\pi^+\pi^-$ contributions to the
RBC/UKQCD windows and $\amuHVP$,
\begin{align}
[a_\mu^{\rm SD}]^{I=1}_{2\pi}\times 10^{10} &= 15.53(13), \nonumber \\
[a_\mu^{\rm int}]^{I=1}_{2\pi}\times 10^{10} &= 150.3(1.2), \nonumber \\
[a_\mu^{\rm LD}]^{I=1}_{2\pi} \times 10^{10} &= 359.4(2.8), \nonumber \\
[a_\mu^{\rm HVP}]^{I=1}_{2\pi} \times 10^{10}&= 525.2(4.2)
\end{align}
are all significantly larger than the purely-KNT19-based
counterparts given in the third row of Tab.~\ref{tab:unambiguous-modes-I1}.
The shift for $\amuHVP$, for example, is about $+21.5\units$. These
exploratory CMD-3-based $2\pi$ results can be combined with the
KNT19-based contributions from all other exclusive modes and
the resulting modified exclusive-mode sums used to reevaluate the
lqc and s+lqd components of the three windows. The results of this
exercise are given in Tab.~\ref{tab:final-results-lqc-CMD3}
and~\ref{tab:final-results-slqd-CMD3}.

As can be seen in the last row of Tab.~\ref{tab:final-results-lqc-CMD3},
the use of the CMD-3 $2\pi$ data produces larger lqc contributions for
all RBC/UKQCD windows. These larger results, moreover, all agree
very well with lattice-QCD determinations of the same quantities,
as shown in Fig.~\ref{fig:results-lqc-CMD3}. The CMD-3-based
result for $a_\mu^{\rm HVP}$ would also produce the shift required
to make the data-driven determination of the full $a_\mu$ compatible with
experiment. A recent analysis \cite{Davies:2024pvv} provides
further evidence for the agreement
between the lattice results and the KNT19 data modified with
the results for the two-pion mode
from CMD-3.

In the CMD-3-based s+lqd results of Tab.~\ref{tab:final-results-slqd-CMD3},
we see that the shifts for the SD and intermediate windows are very
small and thus do not spoil the very good agreement observed above
between the lattice determinations of these quantities and the KNT19-based
results of Tab.~\ref{tab:final-results-slqd}, see Fig.~\ref{fig:results-splqd-KNT19}. For the LD s+lqd
component, where a strong cancellation between the unambiguous- and
ambiguous-mode totals is observed, the shift is somewhat
larger and the CMD-3-based result is $2.96(79)\units$. This shift reduces the difference between the only lattice determination of this quantity, given in Eq.~(\ref{eq:Mainz-LD-splqd}), and the data-driven result from $1.3\sigma$ to $0.7\sigma$. Finally, the result for
$a_\mu^{\rm HVP,s+lqd}$, which is shifted from $40.7(1.7)\units$ to
$38.4(1.7)\units$, remains compatible with lattice-QCD determinations.

\section{Conclusions}
\label{sec:conclusions}

In this paper, we concluded the data-driven evaluation of the lqc
and s+lqd components of the three RBC/UKQCD windows for $\amuHVP$
providing, for the first time, results for the SD and LD windows.
The method we employed is the same as that used in our previous
determinations of the lqc and s+lqd components of the total
$\amuHVP$~\cite{Boito:2022dry,Boito:2022rkw} and RBC/UKQCD intermediate
window~\cite{Benton:2023dci}, as well as those of the alternate
intermediate window of Ref.~\cite{Aubin:2022hgm} and other windows
proposed in the
literature~\cite{Boito:2022njs,Benton:2023fcv}}. We have shown, in this series of papers, that the data-driven
computation can be reorganized to make possible the extraction of
isospin-symmetric results for these key components of $\amuHVP$,
and hence also detailed, precision comparisons with results
obtained from the lattice.

Our data-driven results for the s+lqd components, obtained using the
(pre-CMD-3) KNT19 data compilation, agree well with lattice
determinations, with no significant discrepancies within our
uncertainties. In contrast, results for the data-driven lqc
components of the intermediate and LD windows obtained from the
KNT19 data compilation, summarized in Tab.~\ref{tab:final-results-lqc}
and Fig.~\ref{fig:results-lqc-KNT19}, show a strong tension with
lattice results. In the case of the intermediate window, there
is a greater-than-5$\sigma$ incompatibility between our
data-driven determination and the most precise lattice numbers,
while for the LD window a significant discrepancy is found
with respect to the the two recently released, unblinded lattice results.
Results for the SD window are also systematically lower than the
lattice determinations, though, with our larger data-driven
uncertainty, the resulting data-driven versus lattice differences
are of lower significance than those of the LD and intermediate
window cases.

It is important to note that the data-driven results for the lqc
components of the intermediate and LD windows are dominated by
the $\pi^+\pi^-$ contributions, which represent 80\% and 88\%
of the totals, respectively. These results are, therefore, sensitive
to the discrepancies in the different measurements of the
$e^+e^-\to \pi^+\pi^-$ cross sections. The results for the
SD window, in contrast, are less strongly sensitive to
the $\pi^+\pi^-$ contribution and receive an important contribution
from pQCD.

These observations raise the question of the potential impact of
the CMD-3 $2\pi$ data, which we investigated, in an exploratory
framework, in Sec.~\ref{sec:CMD-3-results}. In this preliminary
analysis, we simply replaced the KNT19 $2\pi$ data with CMD-3
results in the region covered by the CMD-3 experiment. The results for
the lqc component obtained using CMD-3 data, summarized in
Tab.~\ref{tab:final-results-lqc-CMD3} and
Fig.~\ref{fig:results-lqc-CMD3}, are in very good agreement with the
lattice determinations. The CMD-3-induced shifts in the s+lqd
components are, of course, a factor of $10$ smaller than those in the
lqc components, and it is thus not surprising that, as was the case
for s+lqd results based on pre-CMD-3 data, the CMD-3-modified s+lqd
results, given in Tab.~\ref{tab:final-results-slqd-CMD3}, remain
compatible with the lattice. In the single case where a larger shift is observed, namely for the s+lqd LD window component, the CMD-3-based result brings the central value into even closer agreement with that of the recent lattice result from Mainz, Eq.~(\ref{eq:Mainz-LD-splqd}). This demonstrates explicitly that shifts
in the $\rho$-peak-region $2\pi$ cross sections, such as those observed
in the CMD-3 data, can produce data-driven lqc components compatible
with the lattice, without disturbing the good agreement observed for the SD, the intermediate window, and the total HVP s+lqd components.

We emphasize that our comparisons between data-driven and
lattice results
are based on employing the KNT19 (and CMD-3) data for the
data-driven RBC/UKQCD
window results.
At present, with the exclusive-mode distributions
and covariances of the DHMZ collaboration~\cite{Davier:2019can,Davier:2017zfy} not
publicly available, it is not possible to carry out these same comparisons
using DHMZ input.
This is in contrast
to the situation for $\amuHVP$, where DHMZ integrated exclusive-mode-contribution
results are publicly available~\cite{Davier:2019can}, and these
allowed us to obtain, in Refs.~\cite{Boito:2022dry,Boito:2022rkw},
DHMZ-exclusive-mode-based data-driven determinations of the lqc
and s+lqd components of $\amuHVP$. With the recent 2024 BMW
correction to their previous $\amuHVP$ EM IB contribution result,
our previous data-driven lqc result~\cite{Boito:2022dry} is now
shifted slightly, to
\beq
a_\mu^{\rm HVP,lqc}\times 10^{10} = 638.9(4.1)
\qquad (\mbox{DHMZ based~\cite{Davier:2019can}}).
\eeq

\section*{Acknowledgements}
\noindent We thank Aida El-Khadra, Simon Kuberski, and Hartmut Wittig for providing results from as yet unpublished lattice QCD analyses. DB’s work was supported by the S\~ao Paulo Research Foundation
(FAPESP) grant No.~2021/06756-6 and by CNPq grant No. 308979/2021-4.
MG is supported by the U.S. Department of Energy, Office of Science,
Office of High Energy Physics, under Award No. DE-SC0013682. AK is supported by The Royal Society (URF$\backslash$R1$\backslash$231503).
KM is supported by a grant from the Natural Sciences and Engineering
Research Council of Canada. SP is supported by the Spanish Ministerio de Ciencia e Innovacion, grants PID2020-112965GB-I00 and PID2023-146142NB-I00,
and by the Departament de Recerca i Universities from Generalitat de Catalunya
to the Grup de Recerca 00649 (Codi: 2021 SGR 00649).
IFAE is partially funded by the CERCA program of the Generalitat de Catalunya.

\bibliographystyle{jhep-modified}
\bibliography{Refs-SD-and-LD-lqc}

\end{document}